\documentclass[final,3p,times]{elsarticle}
\usepackage{amssymb}
\usepackage{lineno}
\usepackage{amsmath}
\usepackage{soul}
\usepackage{etoolbox}
\usepackage{graphicx}
\usepackage{txfonts} 
\usepackage{bbm}
\usepackage[english]{babel}
\usepackage{empheq} 
\usepackage{mathrsfs} 
\usepackage{calc} 
\usepackage{caption} 
\usepackage{subcaption}
\usepackage{enumitem}                                 
\usepackage{amsthm} 
\usepackage{cases}
\usepackage{siunitx}
\usepackage{upgreek}
\DeclareMathAlphabet{\mathdutchcal}{U}{dutchcal}{m}{n}
\usepackage[colorlinks=true]{hyperref}
\usepackage[hyperref]{xcolor}
\usepackage[normalem]{ulem}
\biboptions{authoryear,comma,compress}
\theoremstyle{remark}

\newcommand{\toVect}[1]{{\boldsymbol{#1}}}

\usepackage{appendix}

\DeclareMathOperator{\sign}{sign}

\DeclareSIUnit \uJ  { \micro \joule }
\DeclareSIUnit \uC  { \micro \coulomb }
\newcommand{\um}{~\si{\um}}

\makeatletter
\def\appendixname{Appendix }
\renewcommand\appendix{\par
  \setcounter{section}{0}%
  \setcounter{subsection}{0}%
  \setcounter{equation}{0}
  \gdef\thefigure{\@Alph\c@section.\arabic{figure}}%
  \gdef\thetable{\@Alph\c@section.\arabic{table}}%
  \gdef\thesection{\appendixname~\@Alph\c@section}%
  \@addtoreset{equation}{section}%
  \gdef\theequation{\@Alph\c@section.\arabic{equation}}%
  \addtocontents{toc}{\string\let\string\numberline\string\tmptocnumberline}{}{}
}

\newdimen\appnamewidth
\def\tmptocnumberline#1{%
   \setbox0=\hbox{\appendixname}
   \appnamewidth=\wd0
   \addtolength\appnamewidth{2.5pc}
   \hb@xt@\appnamewidth{#1\hfill}
}
\makeatother

\journal{}

\begin{document}

\begin{frontmatter}
\title{Flexoelectricity causes surface piezoelectric-like effects in dielectrics}
\author[lacan,cimne]{H. Mohammadi}
\author[lacan,epsem]{F. Greco\corref{mycorrespondingauthor}}\cortext[mycorrespondingauthor]{Corresponding author}\ead{francesco.greco@upc.edu}
\author[lacan,georgia]{D. Codony}
\author[lacan,cimne]{I. Arias}

\address[lacan]{Laboratori de C\`{a}lcul Num\`{e}ric (LaC\`{a}N), Universitat Polit\`{e}cnica de Catalunya (UPC), Campus Nord UPC-C2, E-08034 Barcelona, Spain}
\address[cimne]{Centre Internacional de M{\`e}todes Num{\`e}rics en Enginyeria (CIMNE), 08034 Barcelona, Spain}
\address[epsem]{Department of Mining, Industrial and ICT Engineering, Universitat Polit\`{e}cnica de Catalunya (UPC), Campus Manresa, E-08242 Manresa, Spain}
\address[georgia]{College of Engineering, Georgia Institute of Technology, Atlanta, Georgia 30332, USA}
\begin{abstract} 
In this paper we study the surface effects that bulk flexoelectric models in finite samples exhibit. We first show that when the body is infinite, flexoelectric materials do not exhibit electromechanical response under homogeneous loading. However, when the size of the body is finite, due to the symmetry-breaking nature of surfaces, homogeneous loading (mechanical or electrical) can cause an electromechanical response near the surfaces. We obtain closed-form solutions for finite samples under different electromechanical loading conditions, and show that the electromechanical response caused by the bulk flexoelectric effect is reminiscent of surface piezoelectricity, causing boundary layers in certain components of the strains and/or electric fields near the free surfaces. 
\end{abstract}

\end{frontmatter}

\section{Introduction \label{sec_01}}
Flexoelectricity, a linear couplings between strain gradient and electric field (direct flexoelectricity) or between electric field gradient and strain (converse flexoelectricity), is present in all dielectric materials, including crystals, polymers, biomaterials, liquid crystals, etc. \citep{zubko2013flexoelectric, nguyen2013nanoscale, ahmadpoor2015flexoelectricity,WANG2019100570}.
These couplings, which were first predicted theoretically \citep{mashkevich1957electrical} have since been confirmed experimentally. \citet{bursian1968} demonstrated beam bending of non-piezoelectric thin cantilever beams under applied electric bias in closed circuit, an evidence of inverse flexoelectricity producing non-homogeneous deformations in response to an applied homogeneous electric field. This effect has been later used in proof-of-concept flexoelectric MEMS \citep{bhaskar2016flexoelectric}. \citet{ma2001observation,Ma2002} and \citet{cross2006flexoelectric} conducted a series of experiments showing electric fields emerging in cantilever nanobeams under bending and nanopyramids under compression, a testament of the direct flexoelectric effect. 

The flexoelectric response induced by mechanical gradients has been shown to be strong enough to: (i) switch polarization in ferroelectrics, which opens avenues for mechanical writing of ferroelectric memories without any electrical bias \citep{lu2012mechanical}, (ii) change the conductivity of LAO/STO interfaces by purely mechanical means, which can find application in transistors \citep{sharma2015mechanical}, and (iii) provide a charge separation mechanism in non-centrosymmetric materials for photovoltaic applications \citep{yang2018flexo}. Finally, deformation under inhomogeneous electric fields due to converse flexoelectricity has been observed in Piezoresponse Force Microscopy (PFM) \citep{abdollahi2019converse}. 
All these settings have been successfully modeled with a self-consistent two-way coupled electromechanical continuum framework, demonstrating the ability of the model to capture flexoelectric physics \citep{abdollahi2014computational,abdollahi2015a,codony2021mathematical,abdollahi2019converse}.

\begin{figure}[!t]\centering
\minipage{0.5\textwidth}
\includegraphics[scale=0.25]{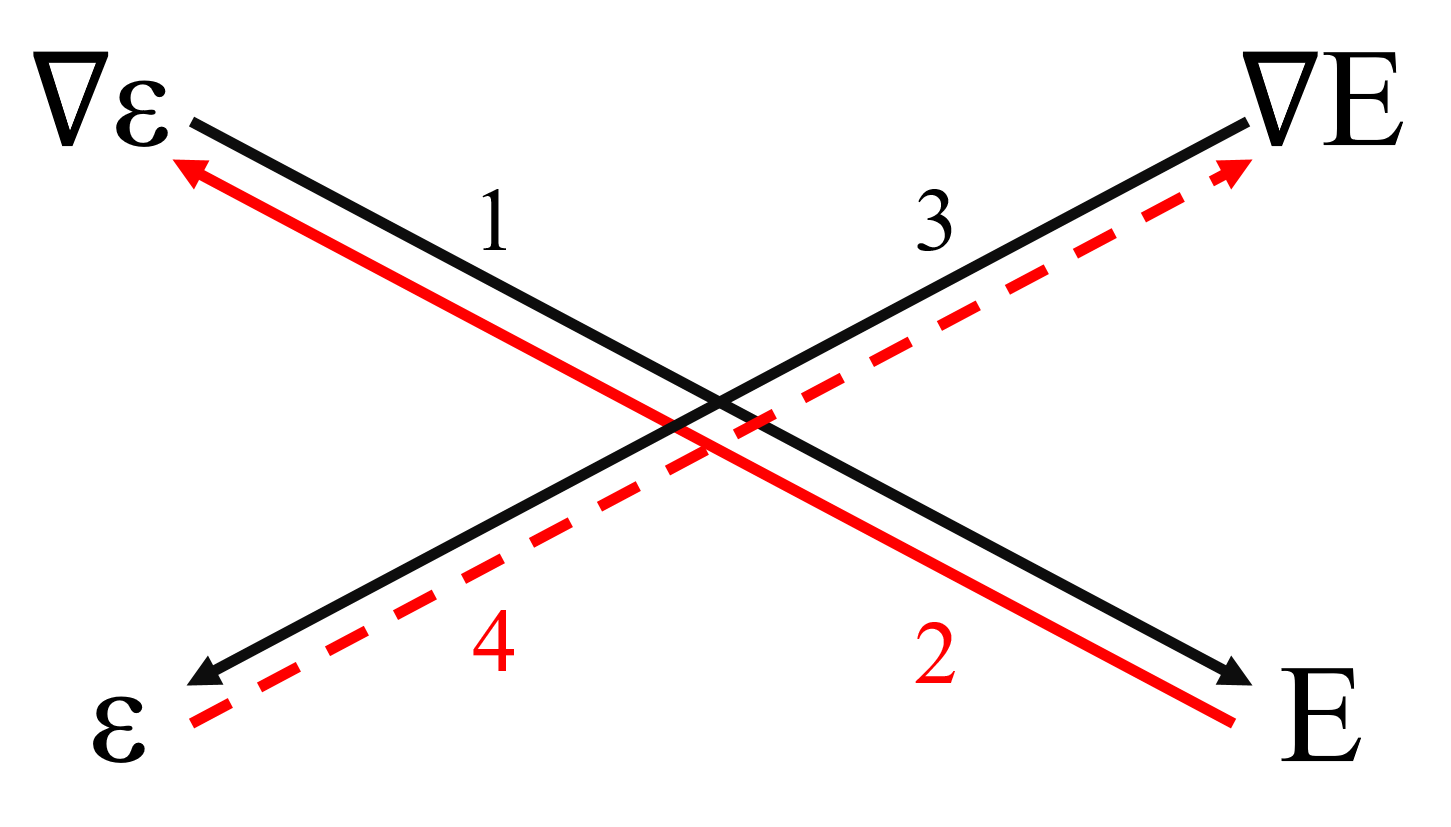}\subcaption{\label{Couplings} Four-way coupling in the continuum models of flexoelectricity, namely (1) direct flexoelectricity, (2) inverse flexoelectricity, (3) converse flexoelectricity, (4) inverse-converse flexoelectricity. The first three phenomena have been confirmed experimentally, while the fourth one has not yet been observed.}
\endminipage\\
\minipage[t]{0.4\textwidth}
\includegraphics[width=\textwidth]{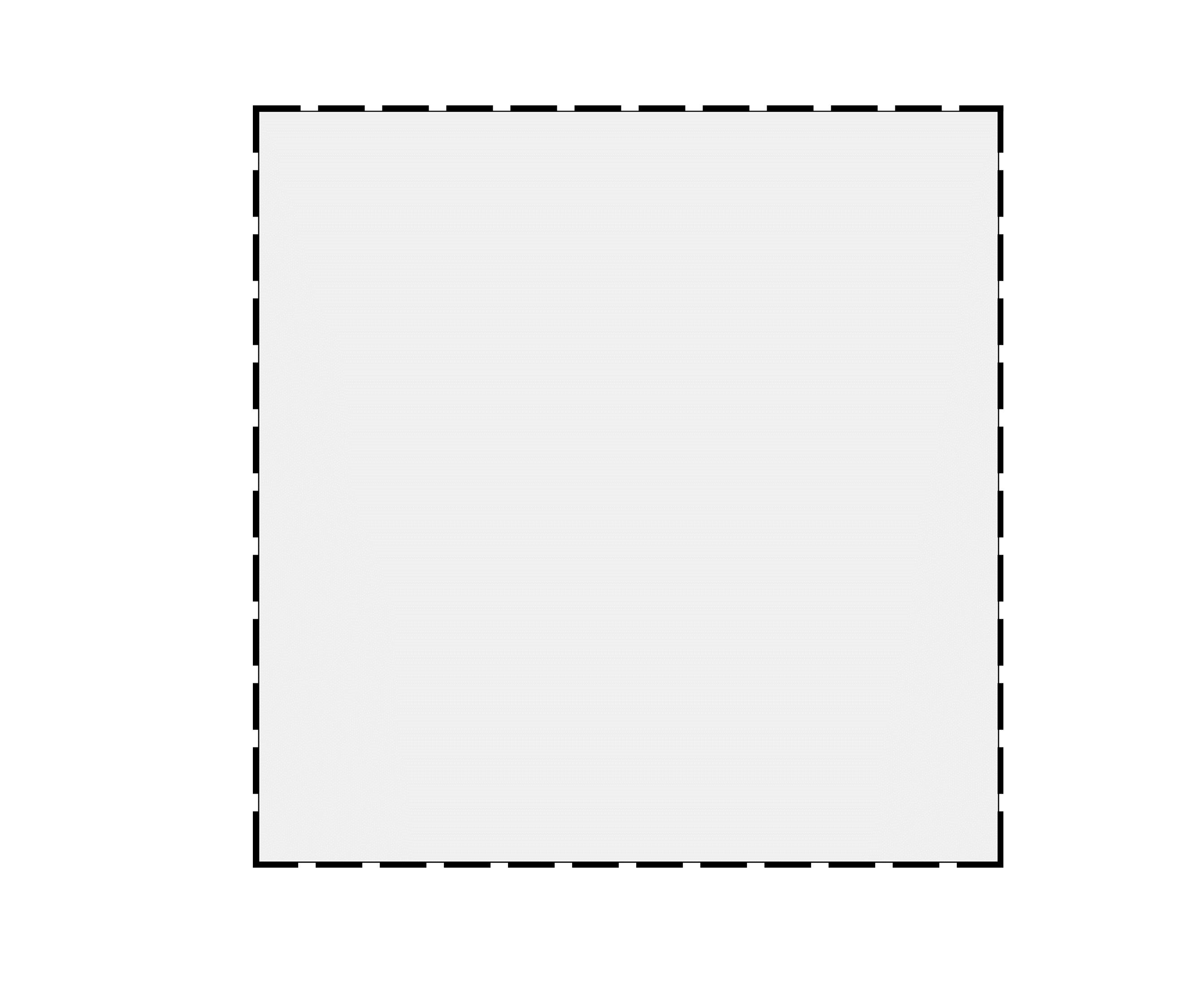}\subcaption{\label{CompressionInfinite} Horizontal ($x$) compression of an infinite flexoelectric body (periodic in $x$ and $y$ directions) does not induce any electric response, as symmetry is not broken with compression.}
\endminipage~
\minipage[t]{0.4\textwidth}
\includegraphics[width=\textwidth]{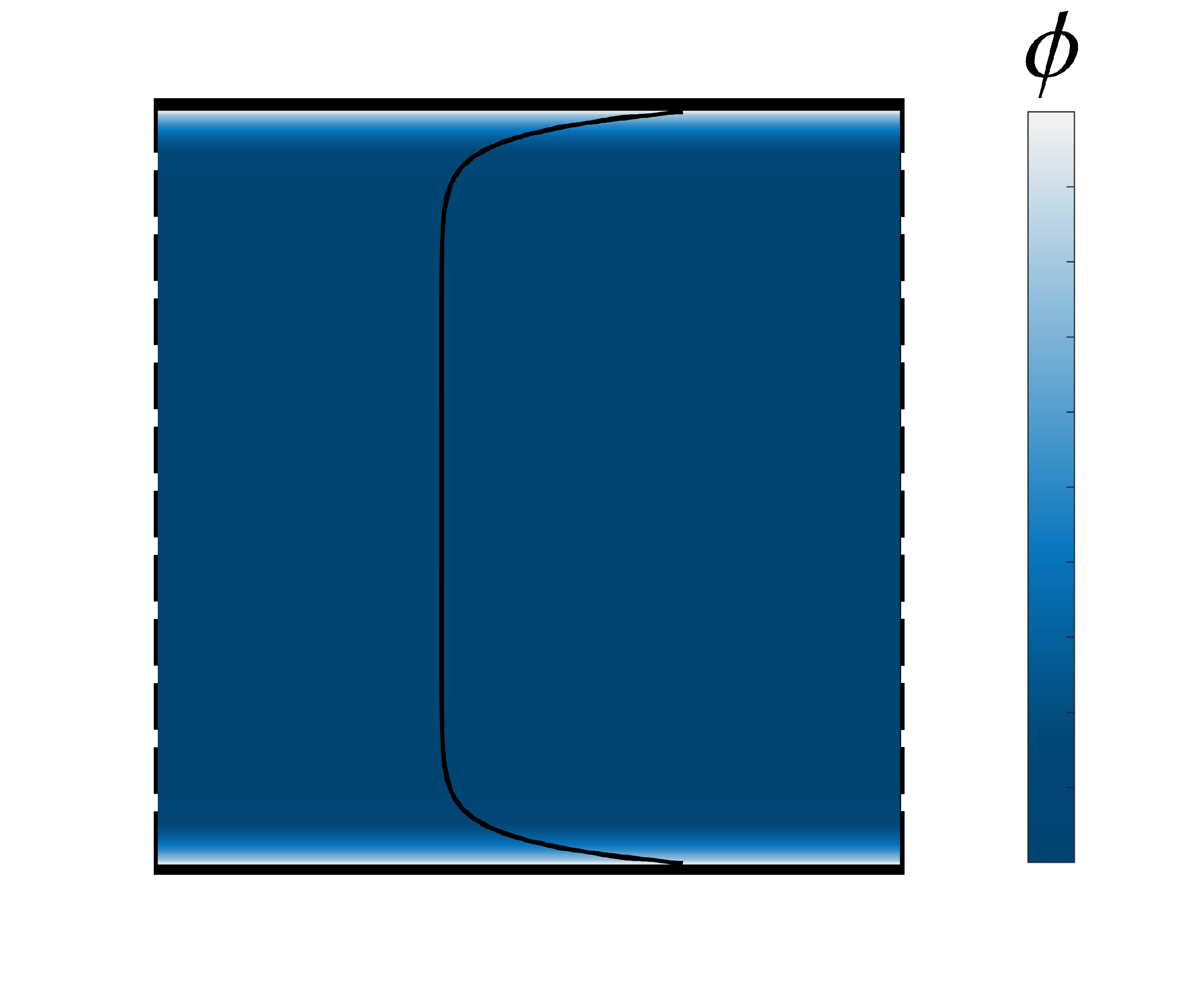}\subcaption{\label{CompressionFinite}  Horizontal ($x$) compression of an infinite flexoelectric film (periodic in $x$ but finite in $y$ direction) induces an electric response close to the free top and bottom boundaries, as surfaces are sources of symmetry-breaking.}
\endminipage
\caption{\label{fig:CompressionBLPeriodic} Flexoelectric couplings. The inverse phenomena (couplings shown with red arrows in (a)) can cause surface effects in flexoelectric models in finite samples. (b) and (c) depict the distribution of electric potential in a Lifshitz-invariant flexoelectric body without/with free surfaces.}
\end{figure}

In all three flexoelectric mechanisms, the actuating field, namely a mechanical gradient (direct), an electric field (inverse) and an electric field gradient (converse) are polar in nature and thus able by themselves to break material centro-symmetry.  This is not expected to happen under homogeneous strain (c.f.~Fig.~\ref{Couplings}). Indeed, the self-consistent simulation of a non-piezoelectric dielectric square sample with generalized periodic boundaries \citep{barcelo2022weak,barcelo2023}, i.e.~representing an infinite medium, under uniform compression shows no flexoelectric response as expected (Fig.~\ref{CompressionInfinite}). Interestingly, simulations in finite samples for  all three models  in Section \ref{sec_02} exhibit a boundary layer in the electric potential or the strain, which vanishes in the bulk as expected (Fig.~\ref{CompressionFinite}). This boundary layer emerges naturally from the model in the presence of a free surface. Intuitively, this localized electric response can be viewed as the piezoelectric-like response of a thin layer of material close to the free surface. It is thus reminiscent of surface piezoelectricity. This effect manifests itself in finite samples as an emerging thin piezoelectric boundary layer resulting from symmetry loss at the surface \citep{zubko2013flexoelectric}.

Regardless of the intrinsic symmetry of the bulk material, the presence of a free surface breaks the symmetry by surface relaxation and induces the emergence of a thin layer of non-centrosymmetric material with piezoelectric-like behavior. Surface piezoelectricity has been modeled as a zero-thickness layer of piezoelectric material, in the spirit of \citet{Tagantsev2012,Yudin2013,yurkov2016strong}. However, it is known that surface relaxation can be described by strain gradient elasticity \citep{danescu2012hyper}.  Similarly, here a piezoelectric-like boundary layer emerges naturally from the rich continuum model, without a specific ad-hoc model for surface piezoelectricity. Similar to the boundary layers in strain gradient elasticity models \citep{lam2003experiments,shu1999finite,shodja2012calculation}, the observed boundary layers present an exponential growth near the surfaces, and their width is directly related to the length scale parameters of the inherent higher-order physics. 

An in-depth understanding of the inherent surface effects of flexoelectric models is essential from modeling, computational, and physical perspectives. From the modeling side, the emergence of boundary layers from surface relaxation in the flexoelectric models needs to be taken into account when incorporating ad-hoc surface piezoelectricity models such as the zero-thickness piezoelectric surface layer as done in \citet{dai2011surface} and \citet{pan2011continuum}. Furthermore, a rigorous characterization of the boundary layers provides an opportunity to model surface effects resulting from surface relaxation as an emergent property. Obviously, surface effects resulting from physical or chemical surface specificity cannot be captured by the present rich continuum models. 
On the computational side \citep{codony2019immersed,yvonnet2017numerical}, the inherent surface effects of the flexoelectric models can cause steep boundary layers resulting in numerical instabilities if the computational mesh is not sufficiently fine. Quantitative knowledge of the inherent surface effects of the flexoelectric models can be a useful guide for careful consideration of the mesh size and/or regularization parameters. Finally, the detailed study of specific boundary value problems based on the rich continuum models can provide further insights on the physics of the free surface effects.

In the following sections, we present a theoretical exploration and characterization of the observed boundary layers. 
We first study two examples in which a homogeneous electric field or strain causes surface effects in a thin flexoelectric film. As an additional example, we then explore the uniform bending of a flexoelectric beam, showing that it exhibits surface effects that could be well-explained with the similarity to those seen due to the application of homogeneous strain. 

\section{Methodology}\label{sec_02} 
We consider a general form of flexoelectric coupling from which different forms such as direct, converse, and Lifshitz-invariant flexoelectric models \citep{codony2021mathematical} can be derived. In the limit of infinitesimal deformations, the electromechanical enthalpy density can be written in terms of strains $\varepsilon_{ij}$, electric fields $E_l$, and their gradients as:
\begin{align}\label{enthalpy}
\mathcal{H}\big(\varepsilon_{ij}, \varepsilon_{ij,k}, E_l, E_{l,m}\big)=& \frac{1}{2} \varmathbb{C}_{ijkl}\varepsilon_{ij}\varepsilon_{kl}+\frac{1}{2} h_{ijklmn}\varepsilon_{ij,k}\varepsilon_{lm,n}+\zeta \mu_{lijk}\varepsilon_{ij}E_{l,k}-(1-\zeta) \mu_{lijk}\varepsilon_{ij,k}E_{l}-\frac{1}{2} \kappa_{lm}E_lE_m-\frac{1}{2} M_{ijkl}E_{i,j}E_{k,l},
\end{align}
where $\zeta=0$ is related to the direct model, $\zeta=1$ is related to the converse model, and  $\zeta=0.5$ is related to the Lifshitz-invariant model. In Eq.~\eqref{enthalpy}, $\toVect{\varmathbb{C}}$ is the elasticity tensor, $\toVect{h}$ is the strain gradient elasticity tensor, $\toVect\mu$ is the flexoelectricity tensor,  $\toVect\kappa$ is the dielectricity tensor, and $\toVect{M}$ is the gradient dielectricity tensor. 
The material tensors are defined in \ref{AppMAT}.
The constitutive equations are:

\begin {align}
\hat\sigma_{ij} &= \frac{\partial \mathcal{H} }{\partial\varepsilon_{ij}} =\varmathbb{C}_{ijkl}\varepsilon_{kl}+\zeta \mu_{lijk}E_{l,k},\\
\widetilde\sigma_{ijk} &= \frac{\partial \mathcal{H} }{\partial\varepsilon_{ij,k}} = h_{ijklmn}\varepsilon_{lm,n}- (1-\zeta) \mu_{lijk}E_{l},\\
\hat{D}_l &=- \frac{\partial \mathcal{H} }{\partial E_{l}}=\kappa_{lm}E_m+(1-\zeta) \mu_{lijk}\varepsilon_{ij,k} ,\\
\widetilde{D}_{ij} &=- \frac{\partial \mathcal{H} }{\partial E_{i,j}}=M_{ijkl}E_{k,l}-\zeta \mu_{lijk}\varepsilon_{kl}.
\end{align}

The physical stress and physical electric displacement are:

\begin{align}\label{Sigma}
\sigma_{ij}=&\frac{\partial\mathcal{H}}{\partial\varepsilon_{ij}}-\bigg(\frac{\partial\mathcal{H}}{\partial\varepsilon_{ij,k}}\bigg)_{,k}=\varmathbb{C}_{ijkl}\varepsilon_{kl}+\mu_{lijk}E_{l,k}-h_{ijklmn}\varepsilon_{lm,nk},
\end{align}
and 
\begin{align}\label{D}
D_{i}=&\frac{\partial\mathcal{H}}{\partial E_{i}}-\bigg(\frac{\partial\mathcal{H}}{\partial E_{i,j}}\bigg)_{,j}= \kappa_{ij}E_j+\mu_{ijkl}\varepsilon_{jk,l}-M_{ijkl}E_{k,lj}.
\end{align}
The strong form of the problem can be written as:
\begin{equation}
\begin{cases}
&\sigma_{ij,j}+f_i^{ext}=0 \quad \text{in } \Omega,\\
&D_{l,l}-q=0\quad \text{in } \Omega,
\end{cases}
\end{equation}
where $f_i^{ext}$ is the external body forces per unit volume, and $q$ represents the external electric free charges per unit volume. The strong form is complemented with the following Neumann boundary conditions \citep{codony2021mathematical}:
\begin{align}
&\left(\hat\sigma_{ij}-\widetilde\sigma_{ijk,k}+\nabla^S_l(n_l)\widetilde\sigma_{ijk}n_k\right)n_j-\nabla^S_j(\widetilde\sigma_{ijk}n_k)=t_i  \quad \text{on } \partial\Omega_t,\label{NBC1}\\
&\widetilde\sigma_{ijk}n_jn_k=r_i \quad \text{on } \partial\Omega_r,\label{NBC2}\\
&-\left(\hat D_l-\widetilde D_{lk,k}+\nabla^S_i(n_i)\widetilde D_{lk}n_k\right)n_l+\nabla^S_l(\widetilde D_{lk}n_k)=w \quad \text{on } \partial\Omega_w,\label{NBC3}\\
&-\widetilde D_{jk}n_jn_k=v \quad \text{on } \partial\Omega_v,\label{NBC4}
\end{align}
where $\toVect n$ is the normal vector to the surface, $\nabla^S$ denotes the surface divergence operator, $\toVect{t}$ is traction, $\toVect{r}$ is double traction, $\toVect{w}$ is surface charge density, and $\toVect{v}$ is double charge density. As discussed in \citet{codony2021mathematical}, in regions where the boundary is not smooth, some additional boundary conditions arise. However, this is not relevant to this study.

\section{Analytical solutions for different cases}
\begin{figure}[!h]\centering
\minipage{.55\textwidth}
\includegraphics[scale=0.3]{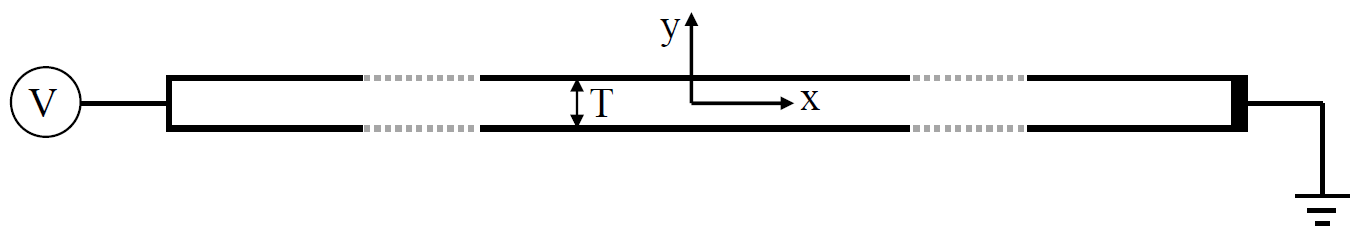}\subcaption{\label{Case1} Electrical actuation of a thin film along its length.}
\endminipage\\
\minipage{.55\textwidth}
\includegraphics[scale=0.3]{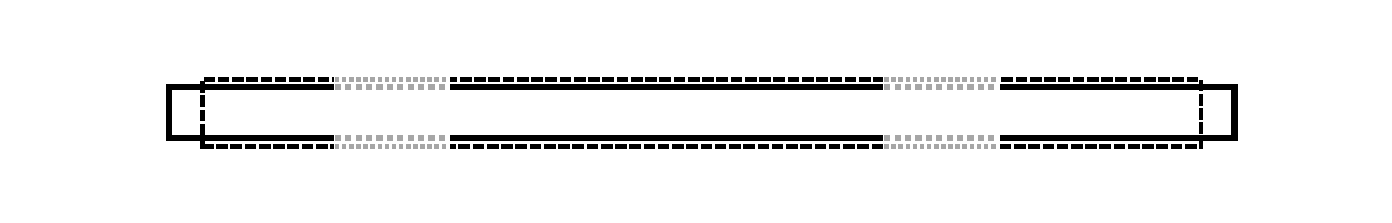}\subcaption{\label{Case2} In-plane compression of a thin film along its length.}
\endminipage\\
\minipage{.55\textwidth}
\includegraphics[scale=0.3]{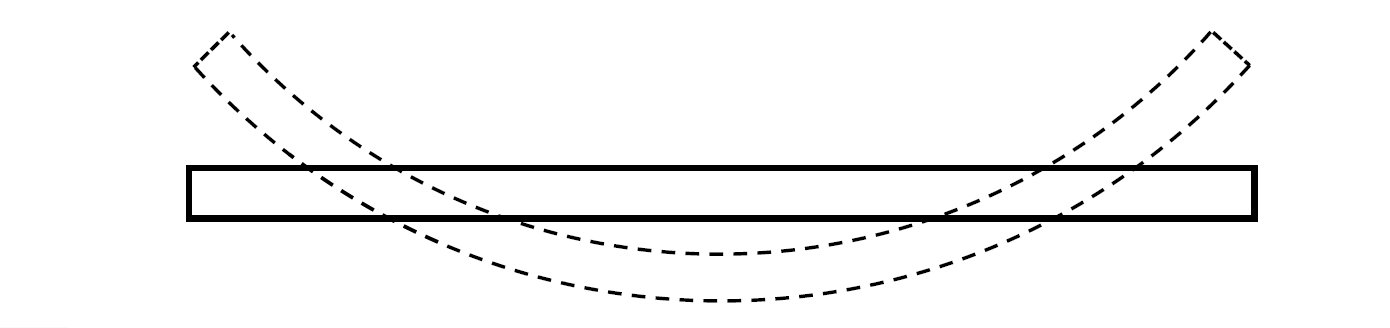}\subcaption{\label{Case3} Uniform bending of a thin film.}
\endminipage
\caption{\label{fig:CasesLoading} Loadings and boundary conditions of three cases studied in the paper. Dark dashed lines represent the deformed shape.}
\end{figure}

We provide next the analytical solution for three boundary value problems corresponding to a plane-strain microfilm under three loading conditions: axial electric actuation (Fig.~\ref{Case1}), axial compression (Fig.~\ref{Case2}), and uniform bending (Fig.~\ref{Case3}). 
For all the examples, we consider a thin flexoelectric film along the $x$-direction occupying $[-T/2, T/2]$ in the $y$-coordinate. The thin film is modeled as being infinite along $x$ and $z$ directions, for one-dimensional kinematics. For the described geometry, the boundary conditions presented in Eqs.~\eqref{NBC1}-\eqref{NBC4} simplify to:
\begin{align}
&t_i = \sigma_{i2} \sign(y) \quad \text{~ ~  ~on } {y=\pm T/2},\label{SNBC1}\\
&r_i = \widetilde{\sigma}_{i22} \qquad~\qquad \text{~ ~on }{y=\pm T/2},\label{SNBC2}\\
&w   = -D_2  \sign(y)   \quad \text{ ~  on } {y=\pm T/2},\label{SNBC3}\\
&v   = -\widetilde{D}_{22}  \qquad~\quad \text{  ~ ~~ on } {y=\pm T/2}.\label{SNBC4}
\end{align}
Homogeneous Neumann mechanical and electric boundary conditions have been considered on the free surfaces for all the cases, i.e. $t_i=0$, $r_i=0$, $w=0$, and $v=0$. The obtained analytical results have been illustrated for each case considering a BST microfilm of thickness $T=1$ micrometer. The material properties are given in Table~\ref{tabcp}. Worth noting that all analytical results have been verified against numerical simulations.

\begin{table}[h]\label{tabmat}
	\centering
		\caption{BST material parameters used in this study. }
	\label{tabcp}
\begin{tabular}[t]{ c| c| c| c|c|c|c|c  }
	\hline
 	E  & $\nu$ & $l_{1}$ & $\epsilon$  & $l_{2}$ & $\mu_L$ & $\mu_T$ &$\mu_S$ \\  
	 \; [Gpa] &   & [nm]&[nC/Vm] &   [nm] &  [$\mu$C/m] & [$\mu$C/m] & [$\mu$C/m] \\ \hline
 	152 & 0.33 & 20 & 8 & 30 & 1.21 & 1.10 & 0.055\\ \hline
\end{tabular}
\end{table}

\subsection{Axial electric actuation \label{BoundaryLayer_AppliedElectricField} }
Suppose we apply a far-field horizontal electric field $\bar E_x$ to the system. Fig.~\ref{Case1} shows the boundary conditions and loadings for this case. 
Consequently, the applied electric field results in $y$-dependent shear strain $\varepsilon_{xy}(y)=\varepsilon_{yx}(y)$. Here we consider strain-free conditions at infinity so that $\varepsilon_{xx}=\varepsilon_{yy}=0$ and $E_y=0$. However, as shown in \ref{App_Case1}, the stress $\sigma_{xx}=0$ on each cross-section which implies that the conclusions are not affected by the choice of stress or strain-free boundary conditions at infinity. Therefore:
\begin{align}
\toVect\varepsilon = 
\begin{bmatrix}
0& \varepsilon_{xy}(y) \\
\varepsilon_{xy}(y)& 0
\end{bmatrix}
,&&
\toVect E= 
\begin{bmatrix}
\bar E_x\\
0
\end{bmatrix}.
\end{align}
Considering homogeneous Neumann boundary conditions on the free surfaces, $\varepsilon_{xy}(y)$ can be obtained as:
\begin{align}\label{Strain_XY}
\varepsilon_{xy}(y)=\frac{-(1-\zeta)\mu_S\bar E_x}{\left(1+\exp(-T/l_1)\right)C_Sl_1}\left[\exp\left(\frac{-y-T/2}{l_1}\right)-\exp\left(\frac{y-T/2}{l_1}\right)\right].
\end{align}
The details of the derivation of the solution are provided in \ref{App_Case1}.
\begin{figure}[!h]\centering
\minipage{0.6\textwidth}
\includegraphics[scale=0.44]{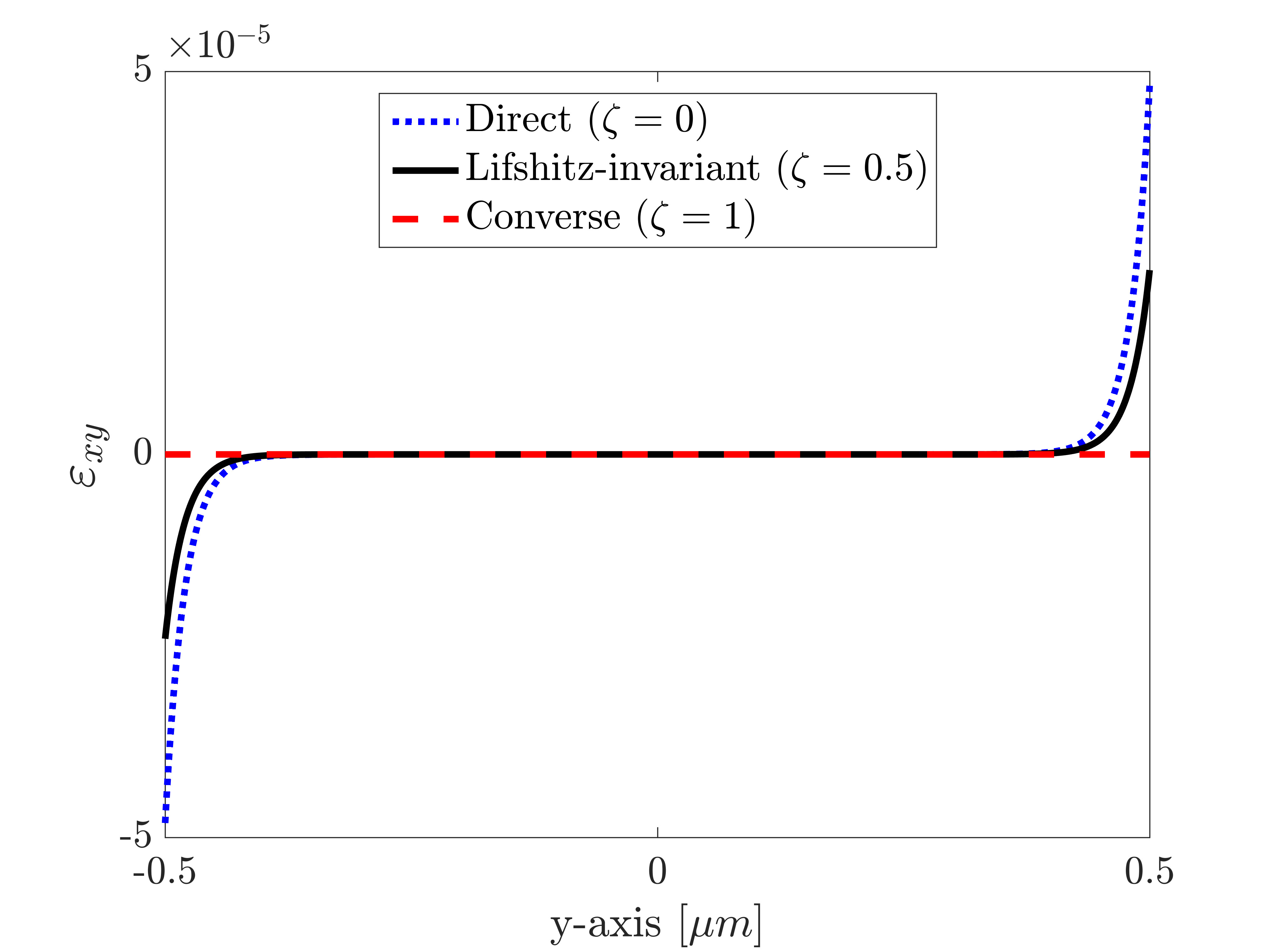}
\endminipage
\caption{\label{fig:ActuationBL_Strain_XY} Electrical actuation of a thin flexoelectric film made of BST along its length with $\bar E_x=1 V/\mu m$ shows inverse surface piezoelectric-like effect.}
\end{figure}

Fig.~\ref{fig:ActuationBL_Strain_XY} depicts $\varepsilon_{xy}$ along the cross-section, Eq.~\eqref{Strain_XY}, for different flexoelectric models. 
It shows that for direct and Lifshitz-invariant models, as a result of the applied in-plane electric field, the top and bottom surfaces of the body experience shear strains. The shear strain vanishes at a certain distance from the surfaces.
The profile of the shear strain is controlled by the strain gradient elasticity lengthscale $l_1$ and its magnitude is proportional to the applied electric field $\bar E_x$ and the shear flexoelectric coefficient $\mu_S$, and is inversely proportional to the shear component of the elasticity tensor $C_S$, and the strain gradient elasticity lengthscale $l_1$. This behavior is similar to the actuation of a thin layer of a piezoelectric surface due to an application of external voltage, or inverse surface piezoelectricity. Fig.~\ref{fig:ActuationBL_Strain_XY} shows that for this case, the converse model does not show surface effect. 

\subsection{Axial compression\label{BoundaryLayer_Compression} }
Suppose we apply far-field plane-strain axial compression $\bar\varepsilon_{xx}$ to the body, Fig.~\ref{Case2}. Both the deformation field and electric potential are independent of the $x$-direction. Therefore, the applied compression results in $y$-dependent vertical strain $\varepsilon_{yy}(y)$ and electric field $E_y(y)$. Note that $\varepsilon_{xy}=\varepsilon_{yx}=0$ and $E_x=0$. Therefore:
\begin{align}
\toVect\varepsilon = 
\begin{bmatrix}
\bar\varepsilon_{xx}& 0 \\
0 & \varepsilon_{yy}(y)
\end{bmatrix},
&&
\toVect E = 
\begin{bmatrix}
0\\
E_y(y)
\end{bmatrix}.
\end{align}

Considering homogeneous Neumann boundary conditions on the free surfaces, $E_y(y)$ and $\varepsilon_{yy}(y)$ can be obtained with the following expressions:
\begin{align}
E_y(y)=K\frac{\bar\varepsilon_{xx}(C_L\mu_T - C_T\mu_L)}{\epsilon C_L}\Biggl[&-\beta_2\left[\exp\left(\frac{-y-T/2}{a_1}\right)-\exp\left(\frac{y-T/2}{a_1}\right)\right]\nonumber\\&+\beta_1\left[\exp\left(\frac{-y-T/2}{a_2}\right)-\exp\left(\frac{y-T/2}{a_2}\right)\right]\Biggr],\label{electricfieldY}\\
\varepsilon_{yy}(y)=K\frac{\bar\varepsilon_{xx}(C_L\mu_T - C_T\mu_L)}{\mu_L C_L}\Biggl[&-a_1\alpha_1\beta_2\left[\exp\left(\frac{-y-T/2}{a_1}\right)+\exp\left(\frac{y-T/2}{a_1}\right)\right]\nonumber\\&+a_2\alpha_2\beta_1\left[\exp\left(\frac{-y-T/2}{a_2}\right)+\exp\left(\frac{y-T/2}{a_2}\right)\right]\Biggr]-\frac{C_T}{C_L}\bar\varepsilon_{xx},\label{strainYY}
\end{align}
where
\begin{align}
K=&\frac{\zeta}{a_1\beta_2\gamma_1 - a_2\beta_1\gamma_2},\\
a_1,a_2=&\sqrt{l_1l_2} \sqrt{\dfrac{A}{1\pm\sqrt{1-A^2}}}, \quad 
A=\frac{2l_1l_2}{l_1^2+l_2^2+l_{\mu}^2}\label{ai},\\
l_{\mu}^2=&\frac{\mu_L^2}{C_L\epsilon}\label{lmu},\\
\alpha_i=&1-\frac{l_2^2}{a_i^2}\label{alphai},\\
\beta_i=&\left(1-\exp(-T/a_i)\right)\left(\frac{l_1^2}{l_\mu^2}\alpha_i+1-\zeta\right)\label{betai},\\
\gamma_i=&\left(1+\exp(-T/a_i)\right)\left(\frac{l_2^2}{a_i^2}(1-\zeta)+\zeta\right)\label{gammai}.
\end{align}
The details of the derivation of the solution are provided in \ref{App_Case2}. Note that in Eqs.~\eqref{betai}, and \eqref{gammai}, the thickness dependence of  $\beta_i$ and $\gamma_i$ vanishes if $T\gg a_i$.

\begin{figure}[!h]\centering
\minipage{0.55\textwidth}
\includegraphics[scale=0.42]{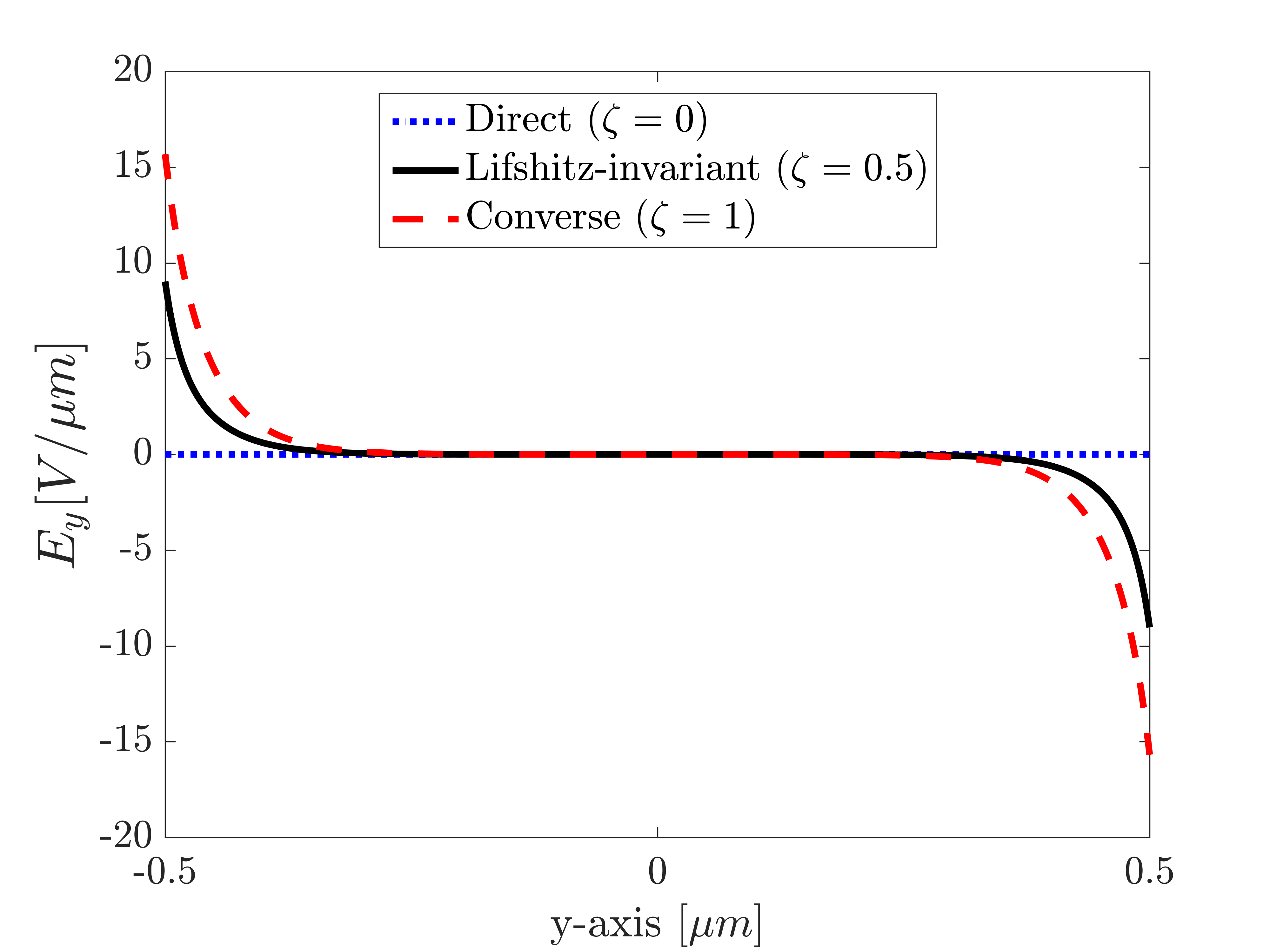}
\endminipage~
\minipage{0.55\textwidth}
\includegraphics[scale=0.42]{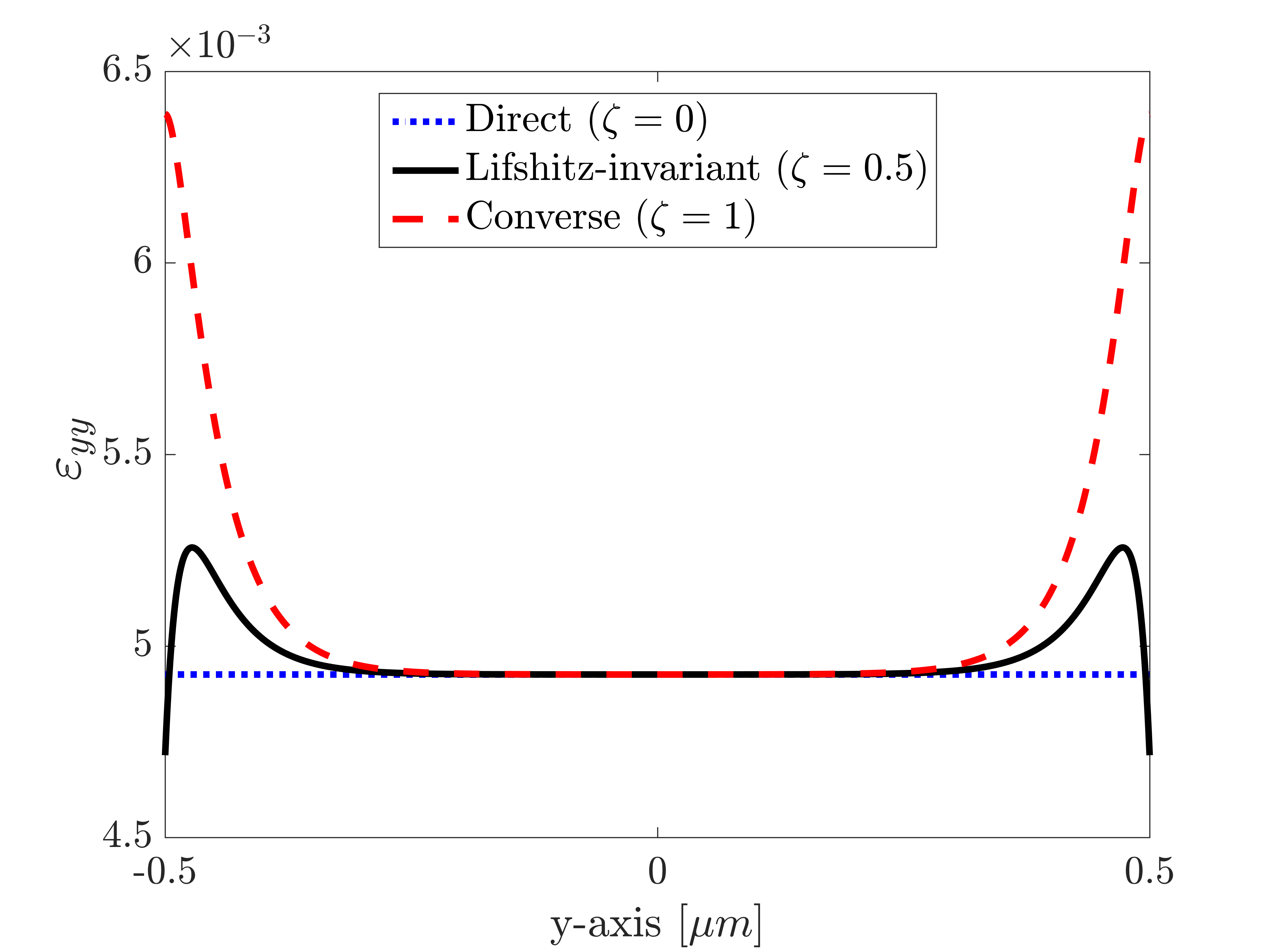}
\endminipage
\caption{\label{fig:CompressionBL} In-plane compression of a thin flexoelectric film made of BST along its length with $\bar\varepsilon_{xx}=-0.01$ shows surface-piezoelectric like effect.}
\end{figure}

Fig.~\ref{fig:CompressionBL} depicts $E_y$ and $\varepsilon_{yy}$ along the cross-section, Eqs.~\eqref{electricfieldY} and \eqref{strainYY}, respectively, for different flexoelectric models. It shows that for Lifshitz-invariant and converse models, a boundary layer develops on the transversal electric field $E_y$, which vanishes away from the surface. This behavior is inherently similar to direct surface piezoelectricity in non-piezoelectric materials, where a thin layer of the surface shows an electric response under mechanical deformation. Due to the generated electric field gradient near the surfaces, the strain $\varepsilon_{yy}$ also experiences a boundary layer due to converse flexoelectricity.

\subsection{Uniform bending\label{BoundaryLayer_Bending}}
Suppose the body is uniformly bent so that its curvature $\kappa$ is the same in all cross-sections normal to it. Fig.~\ref{Case3} shows the boundary conditions and loadings for this case. 
Assuming slenderness, the strains and electric fields can be written as:
\begin{align}
\toVect\varepsilon = 
\begin{bmatrix}
-\kappa y& 0 \\
0 & \varepsilon_{yy}(y)
\end{bmatrix},
&&
\toVect E = 
\begin{bmatrix}
0\\
E_y(y)
\end{bmatrix}.
\end{align}
Considering homogeneous Neumann boundary conditions on the free surfaces, $E_y(y)$ and $\varepsilon_{yy}(y)$ can be obtained with the following expressions:
\begin{align}
E_y(y)=\frac{-\kappa(C_L\mu_T - C_T\mu_L)}{C_L\epsilon}\Biggl[&\hat k_1\left[\exp\left(\frac{-y-T/2}{a_1}\right)+\exp\left(\frac{y-T/2}{a_1}\right)\right]\nonumber\\&-\hat k_2\left[\exp\left(\frac{-y-T/2}{a_2}\right)+\exp\left(\frac{y-T/2}{a_2}\right)\right]\Biggr]-\frac{\kappa}{\epsilon}\left(\mu_L\frac{C_T}{C_L}-\mu_T\right),\label{bendingelectricfield}\\
\varepsilon_{yy}(y)=\frac{-\kappa(C_L\mu_T - C_T\mu_L)}{C_L\mu_L}\Biggl[&\hat k_5\left[\exp\left(\frac{-y-T/2}{a_1}\right)-\exp\left(\frac{y-T/2}{a_1}\right)\right]\nonumber\\&-\hat k_6\left[\exp\left(\frac{-y-T/2}{a_2}\right)-\exp\left(\frac{y-T/2}{a_2}\right)\right]\Biggr]+\frac{C_T}{C_L}\kappa y,\label{bendingstrain}
\end{align}

where
\begin{align}
\hat k_1=&\frac{\left(2a_2\hat\gamma_2(\zeta -1)  + \hat\beta_2T\zeta\right)}{2\left(a_1\hat\beta_2\hat\gamma_1 - a_2\hat\beta_1\hat\gamma_2\right)},\\
\hat k_2=&\frac{ \left(2a_1\hat\gamma_1(\zeta -1)  + \hat\beta_1T\zeta\right)}{2\left(a_1\hat\beta_2\hat\gamma_1 - a_2\hat\beta_1\hat\gamma_2\right)},\\
\hat k_5=&\frac{a_1\alpha_1\left(2a_2\hat\gamma_2(\zeta -1)+ \hat\beta_2T\zeta\right)}{2\left(a_1\hat\beta_2\hat\gamma_1 - a_2\hat\beta_1\hat\gamma_2\right)},\\
\hat k_6=&\frac{ a_2\alpha_2\left(2a_1\hat\gamma_1(\zeta -1) + \hat\beta_1T\zeta\right)}{2\left(a_1\hat\beta_2\hat\gamma_1 - a_2\hat\beta_1\hat\gamma_2\right)},\\
\hat\beta_i=&(1+\exp(-T/a_i))\left(\frac{l_1^2}{l_\mu^2}\alpha_i+1-\zeta\right)\label{hatbetai},\\
\hat\gamma_i=&(1-\exp(-T/a_i))\left(\frac{l_2^2}{a_i^2}(1-\zeta)+\zeta\right)\label{hatgammai}.
\end{align}
where $a_i$, $l_{\mu}$, and $\alpha_i$ have been defined in Eqs.~\eqref{ai}-\eqref{alphai}. The details of the derivation of the solution have been provided in \ref{App_Case3}. Note that in Eqs.~\eqref{hatbetai}, and \eqref{hatgammai}, the thickness dependence of  $\hat\beta_i$ and $\hat\gamma_i$ vanishes if $T\gg a_i$.

\begin{figure}[!h]\centering
\minipage{0.55\textwidth}
\includegraphics[scale=0.42]{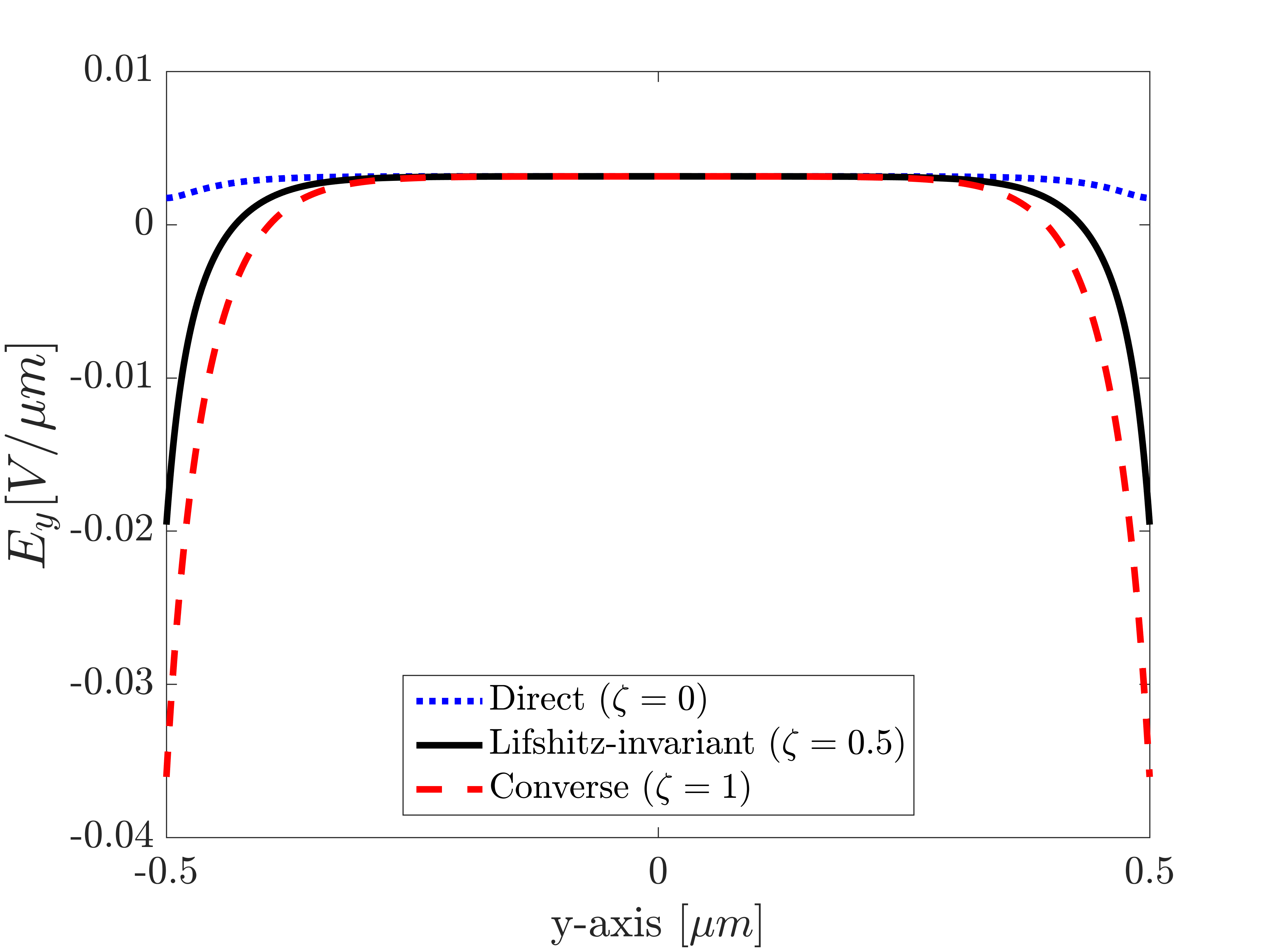}
\endminipage~
\minipage{0.55\textwidth}
\includegraphics[scale=0.42]{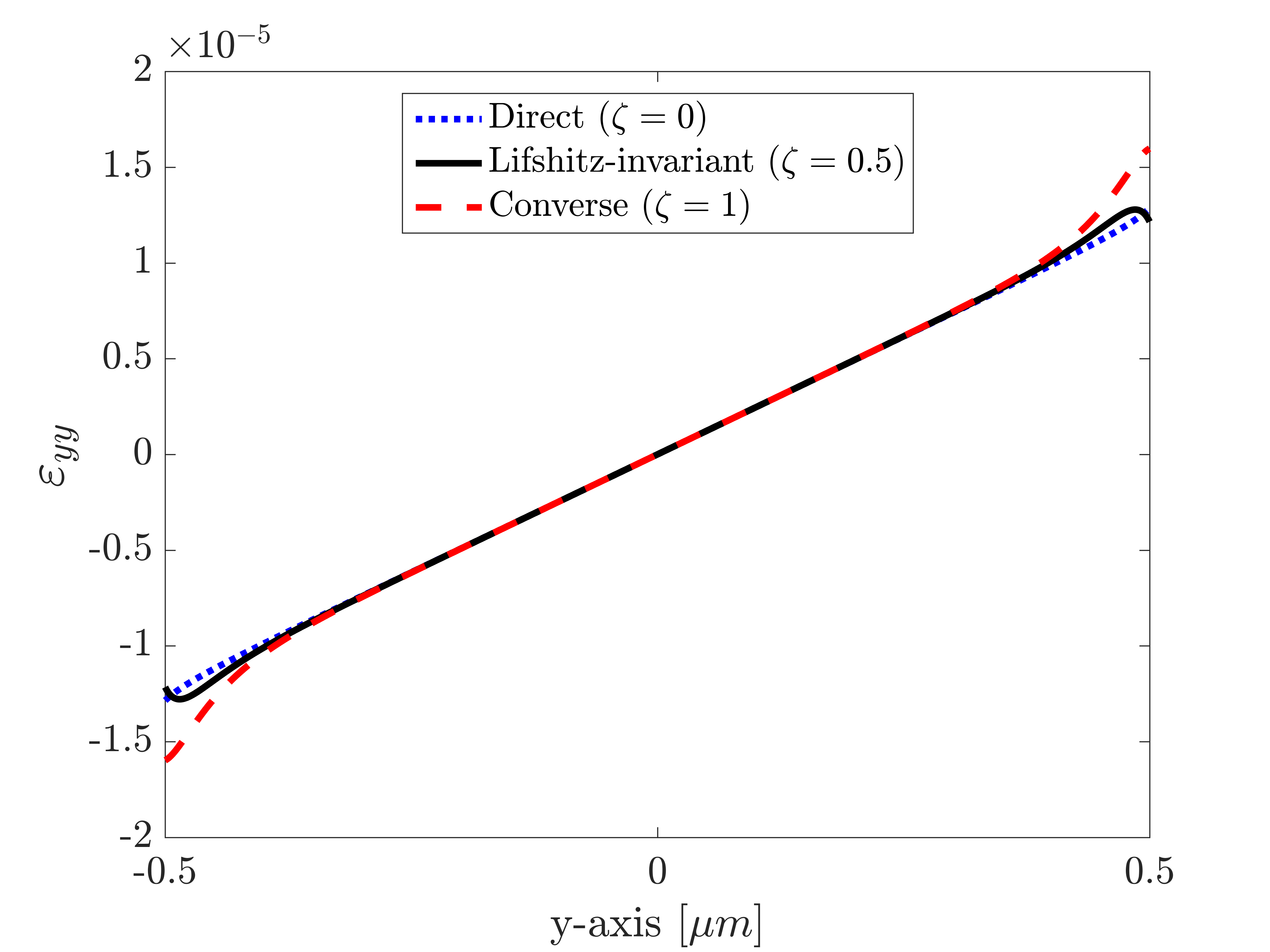}
\endminipage
\caption{\label{fig:BendingBL} Surface effects of a flexoelectric thin film made of BST under uniform bending with $\kappa=50[1/m]$.}
\end{figure}

Fig.~\ref{fig:BendingBL} depicts $E_y$ and $\varepsilon_{yy}$ along the cross-section, Eqs.~\eqref{bendingelectricfield} and \eqref{bendingstrain}, respectively, for different flexoelectric models. It is important to note that the surface effects seen in the case of bending are mainly a combination of the surface effects in the compression case, yet with opposite signs of the applied compression on two sides of the neutral axis. That is why the results shown in Fig.~\ref{fig:BendingBL} exhibit an opposite symmetry compared to Fig.~\ref{fig:CompressionBL}. Furthermore, the coupling between $\varepsilon_{yy,y}$ and $E_y$ through $\mu_L$ causes an additional surface effect. That is why a small surface effect can be seen with the direct model in the case of bending.

\subsection{Discussion}
Three forms of coupling are mainly considered in the flexoelectric literature. Direct flexoelectricity is modeled as $-\mu_{lijk}\epsilon_{ij,k}E_l$, converse flexoelectricity is modeled as $\mu_{lijk}\epsilon_{ij}E_{l,k}$, and Lifshitz-invariant flexoelectricity is modeled as $\frac{1}{2}\mu_{lijk}\epsilon_{ij}E_{l,k}-\frac{1}{2}\mu_{lijk}\epsilon_{ij,k}E_l$. As shown in Section \ref{sec_02}, considering any of the mentioned coupling terms explicitely in the electromechanical enthalpy would not change the governing equations, yet the definition of homogeneous Neumann boundary conditions is different in the three flexoelectric models. This results in solving different boundary value problems if the homogeneous Neumann boundary condition is imposed anywhere on the boundaries. Table \ref{summarize} summarizes the components of the strain and electric field that exhibit boundary layers for the different case studies and different flexoelectric models. Electrical actuation and compression cases are of particular importance as the electromechanical response is isolated from the bulk flexoelectric response. As explained in Section \ref{BoundaryLayer_AppliedElectricField}, a homogeneous electric field can cause a shear strain gradient due to inverse flexoelectricity. This means that the surface of the body exhibits a mechanical response due to the application of an external electrical stimulus, or inverse surface piezoelectricity. Besides, as shown in Section \ref{BoundaryLayer_Compression}, a homogeneous strain can cause an electric field gradient due to inverse converse flexoelectricity, a behavior that is similar to direct surface piezoelectricity.
Considering either the direct or the converse flexoelectric models result in a one-way surface piezoelectric-like effect (direct or inverse), while the Lifshitz-invariant model shows a two-way surface piezoelectric-like effect (direct and inverse). The inherent surface effects of flexoelectricity with different models have already been seen in different studies \citep{yurkov2016strong,abdollahi2014computational,zhuang2020computational,codony2021mathematical}. In particular, \cite{codony2021mathematical} studied a cantilever beam under bending and showed that the Lifshitz-invariant model exhibits a boundary layer on $E_y$. This is in agreement with the results of this paper. However, as in \citet{codony2021mathematical} the longitudinal flexoelectric coefficient $\mu_L$ was neglected, no boundary layer in $E_y$ was observed in the direct model. The boundary layers have also been seen in a cantilever beam actuator \citep{abdollahi2014computational,he2019characterizing,zhuang2020computational,codony2021mathematical}. 

This analysis shed light on the origins of the surface effects in flexoelectric films, which for conciseness, have been considered in open-circuit conditions only. One could extend this analysis to short-circuited films by changing the boundary conditions (applying Dirichlet electric boundary conditions instead of homogeneous Neumann electric boundary conditions on free surfaces) to explain boundary layers present in that case.

\begin{table}
\centering
\caption{\label{summarize}The components of the strain or electric field that exhibit boundary layers for different cases and different flexoelectric models }
\begin{tabular}{ |p{3cm}|p{2cm}|p{3cm}|p{2cm}| }
\hline
Case& Direct &Lifshitz-invariant&Converse\\
~&  ($\zeta=0$) &($\zeta=0.5$)& ($\zeta=1$)\\
\hline
Electric actuation & $\varepsilon_{xy}$ &$\varepsilon_{xy}$ &-\\
Compression & -   & $\varepsilon_{yy},E_y$ &$\varepsilon_{yy},E_y$\\
Bending  &$\varepsilon_{yy},E_y$ & $\varepsilon_{yy},E_y$ &$\varepsilon_{yy},E_y$\\
\hline
\end{tabular}
\end{table}

\subsection{Conclusions}
In this work, we explored the continuum models of flexoelectricity in dielectrics. We showed that when the size of the body is finite, the continuum models of flexoelectricity in bulk exhibit surface piezoelectric-like effects. We attributed the surface effects to inverse flexoelectricity and inverse converse flexoelectricity. We showed that the direct and converse flexoelectric models exhibit a one-way surface piezoelectric-like effect, while the Lifshitz-invariant model shows a two-way surface piezoelectric-like effect. Furthermore, we characterized the observed boundary layers in terms of the lengthscales of the model. These findings may be critical in the design of sound one and two-dimensional analytical models for flexoelectric beams and films, which otherwise assume no boundary layers across the reduced dimension \citep{yan2011surface,yan2013flexoelectric,codony2020modeling,CHU2020105282,wang2024free,mishra2025modeling,shang2022flexoelectricity}, and  may contribute overall in applications with an high surface to volume ratio, such as the development of lattice models for flexoelectric-metamaterials \citep{barcelo2024computational,greco2024topology}.

Future research can be carried out to explore the interaction of the inherent surface effects of flexoelectricity with other physics (bulk piezoelectricity, surface piezoelectricity, and surface flexoelectricity). Besides, how to model flexoelectricity is still an open question (there is no clear physical understanding that which $\zeta$ parameter can well describe reality). There are also no clear insights into the high-order boundary conditions for which homogeneous Neumann conditions are commonly imposed in the literature for convenience. Furthermore, characterization of the physical lengthscales of the model still needs to be further researched. By comparing with other approaches (atomistic simulations or experiments), the results provided here may be useful to find some of the parameters that are not yet well understood.

\section*{Acknowledgments}
This work was supported by the Generalitat de Catalunya (“ICREA Academia” award for excellence in research to I.A., and Grants No.~2017-SGR-1278 and 2021-SGR-01049), the European Research Council (StG-679451 to I.A.), and the Spanish Ministry of Economy and Competitiveness (Grant PID2023-152533OB-I00 funded by MICIU/AEI/10.13039/501100011033/ERDF/EU,
and Grant RTI2018-101662-B-I00). D.C. acknowledges the support of the Spanish Ministry of Universities through the Margarita Salas fellowship (European Union - NextGenerationEU). H.M.~received the support of a fellowship from “la Caixa” Foundation (ID 100010434). The fellowship code is LCF/BQ/DI20/11780036. CIMNE is recipient of a Severo Ochoa Award of Excellence from the MINECO.

\appendix

\section{Material characterization}\label{AppMAT}
The electromechanical enthalpy density described in Eq.~\eqref{enthalpy} involves five material tensors. We define them in this appendix. Below, the non-specified components of the material tensors are zero.


Isotropic elasticity tensor is considered here as
\begin{align}
\varmathbb{C}_{iiii}&=C_L,&&& i&=1,2,\nonumber\\
\varmathbb{C}_{iijj}&=C_T,&&& i,j&=1,2\ \text{with} \ i\neq j,\nonumber\\
\varmathbb{C}_{ijij}=\varmathbb{C}_{ijji}&=C_S,&&& i,j&=1,2\ \text{with} \ i\neq j.
\end{align}
In plain strain condition, $C_L$,$C_T$ and $C_S$ are defined in terms of elasticity modulus $Y$ and Poisson's ratio $\nu$ as
\begin{align}\label{eq_c3}
C_L\coloneqq\frac{Y\left(1-\nu\right)}{(1+\nu)(1-2\nu)},&&
C_T\coloneqq\frac{Y\nu}{(1+\nu)(1-2\nu)},&&
C_S=\frac{C_L-C_T}{2}\coloneqq\frac{Y}{2(1+\nu)}.
\end{align}

We use a simplified form of isotropic strain elasticity tensor which depends on the elasticity modulus $Y$, the Poisson ratio $\nu$ and the internal length scale $l_1$ as \citep{mindlin1964micro,altan1997some}
\begin{align}
	h_{iikiik}&=l_1^2C_L,&&& i,k&=1,2,\nonumber\\
	h_{iikjjk}&=l_1^2C_T,&&& i,j,k&=1,2\ \text{with} \  i\neq j,\nonumber\\
	h_{ijkijk}=h_{ijkjik}&=l_1^2C_S,&&&i,j,k&=1,2\ \text{with} \  i\neq j.
\end{align}
where the parameters $C_L$, $C_S$ and $C_T$ are defined in Eq.~\eqref{eq_c3}.

We use a second-order tensor to describe isotropic dielectricity $\toVect{\kappa}$, which depends on the electric permittivity $\epsilon$ as 
\begin{align}
	\kappa_{ii}&=\epsilon,&&& i&=1,2.
\end{align}

Flexoelectricity is represented by a fourth-order tensor $\toVect{\mu}$. Ferroelectric perovskites in their paraelectric phase are characterized by a cubic-symmetric flexoelectricity tensor involving only three independent flexoelectric coefficients, namely longitudinal $\mu_\text{L}$, transverse $\mu_\text{T}$ and
shear $\mu_\text{S}$. 
\begin{align}
	{\mu}_{iiii}&=\mu_L,&&& i&=1,2,\nonumber\\
	{\mu}_{ijji}&=\mu_T,&&& i,j&=1,2\ \text{with} \  i\neq j,\nonumber\\
    {\mu}_{iijj}={\mu}_{ijij}&=\mu_S,&&& i,j&=1,2\ \text{with} \  i\neq j.
\end{align}

We consider a simplified isotropic gradient dielectricity tensor which depends on electric permittivity of the material $\epsilon$ and the length scale $\ell_2$ in the following form \citep{Mindlin1968b}:
\begin{align}\label{grdieletensor}
M_{ijij}&=\epsilon\ell_2^2,&&& i,j&=1,2.
\end{align}

\section{Analytical derivation for the case of electrical actuation of a thin flexoelectric film along its length\label{App_Case1} }
Consider a thin flexoelectric film along the $x$-direction occupying $[-T/2, T/2]$ in the $y$-coordinate. Suppose we apply far-field horizontal electric field $\bar E_x$ to the system. Consequently, the applied electric field results in $y$-dependent shear strain $\varepsilon_{xy}(y)=\varepsilon_{yx}(y)$. Note that $\varepsilon_{xx}=\varepsilon_{yy}=0$ and $E_y=0$. Therefore:
\begin{align}
\toVect\varepsilon = 
\begin{bmatrix}
0& \varepsilon_{xy}(y) \\
\varepsilon_{xy}(y)& 0
\end{bmatrix},
&&
\toVect E = 
\begin{bmatrix}
\bar E_x\\
0
\end{bmatrix}.
\end{align}

Accordingly, components of the constitutive equations can be written as:
\begin {align}
\hat\sigma_{xy} &=\hat\sigma_{yx} =2C_S\varepsilon_{xy}(y),\\
\hat\sigma_{xx} &=\hat\sigma_{yy}=0,\\~\nonumber\\
\widetilde\sigma_{xyy}&=\widetilde\sigma_{yxy}=2C_Sl_1^2\varepsilon_{xy,y}-2(1-\zeta)\mu_S\bar E_x ,\\
\widetilde\sigma_{xxx}&=-(1-\zeta)\mu_L\bar E_x,\\
\widetilde\sigma_{yyx}&=-(1-\zeta)\mu_T\bar E_x,\\
\widetilde\sigma_{yyy} &=\widetilde\sigma_{xyx} =\widetilde\sigma_{yxx} =\widetilde\sigma_{xxy}=0,\\~\nonumber\\
\hat{D}_x &=\epsilon \bar E_x+2(1-\zeta)\mu_S\varepsilon_{xy,y}(y) ,\\
\hat{D}_y &=0 ,\\~\nonumber\\
\widetilde{D}_{xy} &=\widetilde{D}_{yx}=-2\zeta\mu_S\varepsilon_{xy}(y),\\
\widetilde{D}_{xx} &=\widetilde{D}_{yy} =0.
\end{align}

Therefore, the components of physical stress and physical electric displacement are:
\begin{align}
\sigma_{xy}=&\sigma_{yx}=2C_S\varepsilon_{xy}(y)-2C_Sl_1^2\varepsilon_{xy,yy}(y),\\
\sigma_{xx}=&\sigma_{yy}=0,\\~\nonumber\\
D_{x}=&\epsilon \bar E_x+2\mu_S\varepsilon_{xy,y}(y)\\
D_{y}=&0.
\end{align}

Therefore, the following equilibrium equation in the $y$-direction is the only non-trivial equation that needs to be satisfied:
\begin{align}
2C_S\varepsilon_{xy,y}(y)-2l_1^2C_S\varepsilon_{xy,yyy}(y)=0,\label{equilibrium2ndProblem}
\end{align}
Note that the electrical equilibrium equation and its associated boundary conditions in the $y$-direction are trivially satisfied. The solution of the ODE \eqref{equilibrium2ndProblem} must then satisfy the following non-trivial boundary condition that is the homogeneous Neumann boundary condition related to double traction (Eq.~\eqref{SNBC2}):
\begin{align}
& 2l_1^2C_S\varepsilon_{xy,y}(y)-2(1-\zeta)\mu_S \bar E_{x}=0,  \quad \text{on} \quad  y=\{-T/2, T/2\}.\label{highorderNeumann2ndProblem}
\end{align}
Note that the low-order mechanical boundary condition (Eq.~\eqref{SNBC1}) in the $y$-direction is also trivially satisfied. The following is the solution for the above ODE \eqref{equilibrium2ndProblem} satisfying Eq.~\eqref{highorderNeumann2ndProblem}:
\begin{align}
\varepsilon_{xy}(y)=\frac{-(1-\zeta)\mu_S\bar E_x}{\left(1+\exp(-T/l_1)\right)C_Sl_1}\left[\exp\left(\frac{-y-T/2}{l_1}\right)-\exp\left(\frac{y-T/2}{l_1}\right)\right].
\end{align}

\section{Analytical derivation for the case of in-plane compression of a thin flexoelectric film along its length\label{App_Case2} }
Consider a thin flexoelectric film along the $x$-direction occupying $[-T/2, T/2]$ in the $y$-coordinate. Suppose we apply far-field plane-strain horizontal compression $\bar\varepsilon_{xx}$ to the system. Both the deformation field and electric potential are independent of $x$-direction. Therefore, the applied compression results in $y$-dependent vertical strain $\varepsilon_{yy}(y)$ and electric field $E_y(y)$. Note that $\varepsilon_{xy}=\varepsilon_{yx}=0$ and $E_x=0$. Therefore:
\begin{align}
\toVect\varepsilon = 
\begin{bmatrix}
\bar\varepsilon_{xx}& 0 \\
0 & \varepsilon_{yy}(y)
\end{bmatrix},
&&
\toVect E = 
\begin{bmatrix}
0\\
E_y(y)
\end{bmatrix}.
\end{align}
Accordingly, components of the constitutive equations can be written as:
\begin {align}
\hat\sigma_{xx} &=C_L\bar\varepsilon_{xx}+C_T\varepsilon_{yy}(y)+\zeta \mu_T E_{y,y}(y),\\
\hat\sigma_{yy} &=C_T\bar\varepsilon_{xx}+C_L\varepsilon_{yy}(y)+\zeta\mu_L E_{y,y}(y),\\
\hat\sigma_{xy} &=\hat\sigma_{yx} =0,\\~\nonumber\\
\widetilde\sigma_{yyy} &= l_1^2C_L\varepsilon_{yy,y}(y)- (1-\zeta) \mu_L E_{y}(y),\\
\widetilde\sigma_{xyx} &=\widetilde\sigma_{yxx} =- (1-\zeta) \mu_S E_{y}(y),\\
\widetilde\sigma_{xxy} &=l_1^2C_T\varepsilon_{yy,y}(y)- (1-\zeta) \mu_T E_{y}(y),\\
\widetilde\sigma_{xxx} &=\widetilde\sigma_{xyy}=\widetilde\sigma_{yyx}=\widetilde\sigma_{yxy}=0,\\~\nonumber\\
\hat{D}_y &=\epsilon E_y(y)+(1-\zeta) \mu_L\varepsilon_{yy,y}(y) ,\\
\hat{D}_x &=0 ,\\~\nonumber\\
\widetilde{D}_{xx} &=-\zeta \mu_L\bar\varepsilon_{xx}-\zeta \mu_T\varepsilon_{yy}(y),\\
\widetilde{D}_{yy} &=l_2^2\epsilon E_{y,y}(y)-\zeta \mu_L\varepsilon_{yy}(y)-\zeta \mu_T\bar\varepsilon_{xx},\\
\widetilde{D}_{xy} &=\widetilde{D}_{yx}=0.
\end{align}
Therefore, the components of physical stress and physical electric displacement are:
\begin{align}
\sigma_{xx}=&C_L\bar\varepsilon_{xx}+C_T\varepsilon_{yy}(y)+ \mu_T E_{y,y}(y)-l_1^2C_T\varepsilon_{yy,yy}(y),\\
\sigma_{yy}=&C_T\bar\varepsilon_{xx}+C_L\varepsilon_{yy}(y)+ \mu_L E_{y,y}(y)-l_1^2C_L\varepsilon_{yy,yy}(y),\label{SigmaYY}\\
\sigma_{xy}=&\sigma_{yx}=0,\\~\nonumber\\
D_{y}=& \epsilon E_y(y)+\mu_L\varepsilon_{yy,y}(y)-l_2^2\epsilon E_{y,yy}(y),\label{DY}\\
D_{x}=&0.
\end{align}
Therefore, equilibrium equations in $y$-direction read:
\begin{align}
&C_L\varepsilon_{yy,y}(y)+ \mu_L E_{y,yy}(y)-l_1^2C_L\varepsilon_{yy,yyy}(y)=0,\label{equilibrium1}\\
& \epsilon E_{y,y}(y)+\mu_L\varepsilon_{yy,yy}(y)-l_2^2\epsilon E_{y,yyy}(y)=0\label{equilibrium2},
\end{align}
subjected to high-order homogeneous Neumann boundary conditions (Eqs.~\eqref{SNBC2}, \eqref{SNBC4}):
\begin{align}
& l_1^2C_L\varepsilon_{yy,y}(y)-(1-\zeta) \mu_L E_{y}(y)=0,  \quad \text{on} \quad  y=\{-T/2, T/2\},\label{highorderNeumann1}\\
& l_2^2\epsilon E_{y,y}(y)-\zeta \mu_L\varepsilon_{yy}(y)-\zeta \mu_T\bar\varepsilon_{xx}=0,   \quad \text{on} \quad   y=\{-T/2, T/2\}\label{highorderNeumann2}.
\end{align}
The low-order homogeneous Neumann boundary conditions (Eqs.~\eqref{SNBC1},\eqref{SNBC3}) will naturally be satisfied with the solution we will postulate. 

Extracting $E_{y,y}$ and $\varepsilon_{yy,y}$ from \eqref{SigmaYY} and \eqref{DY}, respectively, and replacing them into \eqref{equilibrium1} and \eqref{equilibrium2}, respectively, we obtain:
\begin{align}
&(l_1^2l_2^2)E_{y,yyyy}(y)-(l_1^2+l_2^2+l_{\mu}^2)E_{y,yy}(y)+E_y(y)=\frac{D_y-l_1^2D_{y,yy}}{\epsilon},\\
&(l_1^2l_2^2)\varepsilon_{yy,yyyy}(y)-(l_1^2+l_2^2+l_{\mu}^2)\varepsilon_{yy,yy}(y)+\varepsilon_{yy}(y)=\frac{\sigma_{yy}-l_2^2\sigma_{yy,yy}}{C_L}-\frac{C_T}{C_L}\bar\varepsilon_{xx}.
\end{align}
where $l_{\mu}^2=\frac{\mu_L^2}{C_L\epsilon}$.

One can postulate the solutions to be of the form:
\begin{align}
E_y=&k_1\exp\left(\frac{-y-T/2}{a_1}\right)+k_2\exp\left(\frac{-y-T/2}{a_2}\right)+k_3\exp\left(\frac{y-T/2}{a_1}\right)+k_4\exp\left(\frac{y-T/2}{a_2}\right),\label{electricfield}\\
\varepsilon_{yy}(y)=&k_5\exp\left(\frac{-y-T/2}{a_1}\right)+k_6\exp\left(\frac{-y-T/2}{a_2}\right)+k_7\exp\left(\frac{y-T/2}{a_1}\right)+k_8\exp\left(\frac{y-T/2}{a_2}\right)-\frac{C_T}{C_L}\bar\varepsilon_{xx},\label{strain}
\end{align}
for $y\in [-T/2, T/2]$, with:
\begin{align}
a_1,a_2 = \sqrt{\dfrac{2l_1^2l_2^2}{l_1^2+l_2^2+l_{\mu}^2\pm\sqrt{(l_1^2+l_2^2+l_{\mu}^2)^2-4l_1^2l_2^2}}}.
\end{align}

Note that $k_1=-k_3$,  $k_2=-k_4$, $k_5=k_7$, $k_6=k_8$ due to symmetry.
Substituting the $E_y$ and $\varepsilon_{yy}$ from Eqs.~\eqref{electricfield} and \eqref{strain} to Eqs.~\eqref{equilibrium1}, \eqref{equilibrium2}, \eqref{highorderNeumann1} and \eqref{highorderNeumann2}, we obtain the unknowns $k_i$. 
\begin{align}
k_1=-k_3=&\frac{-\beta_2\bar\varepsilon_{xx}\zeta(C_L\mu_T - C_T\mu_L)}{C_L\epsilon(a_1\beta_2\gamma_1 - a_2\beta_1\gamma_2)},\\
k_2=-k_4=&\frac{\beta_1\bar\varepsilon_{xx}\zeta(C_L\mu_T - C_T\mu_L)}{C_L\epsilon(a_1\beta_2\gamma_1 - a_2\beta_1\gamma_2)},\\
k_5=k_7=&\frac{-a_1\alpha_1\beta_2\bar\varepsilon_{xx}\zeta(C_L\mu_T - C_T\mu_L)}{C_L\mu_L(a_1\beta_2\gamma_1 - a_2\beta_1\gamma_2)},\\
k_6=k_8=&\frac{a_2\alpha_2\beta_1\bar\varepsilon_{xx}\zeta(C_L\mu_T - C_T\mu_L)}{C_L\mu_L(a_1\beta_2\gamma_1 - a_2\beta_1\gamma_2)},
\end{align}
where
\begin{align}
\alpha_i=&1-\frac{l_2^2}{a_i^2},\\
\beta_i=&\left(1-\exp(-T/a_i)\right)\left(\frac{l_1^2}{l_\mu^2}\alpha_i+1-\zeta\right),\\
\gamma_i=&\left(1+\exp(-T/a_i)\right)\left(\frac{l_2^2}{a_i^2}(1-\zeta)+\zeta\right).
\end{align}

\section{Analytical derivation for the case of uniform bending of a slender beam}\label{App_Case3} 
Consider thin flexoelectric film along the $x$-direction occupying $[-T/2, T/2]$ in the $y$-coordinate. It is uniformly bent so that its curvature $\kappa$ is the same in all cross-sections normal to it.  Assuming slenderness, the strains and electric fields can be written as:
\begin{align}
\toVect\varepsilon = 
\begin{bmatrix}
-\kappa y& 0 \\
0 & \varepsilon_{yy}(y)
\end{bmatrix},
&&
\toVect E = 
\begin{bmatrix}
0\\
E_y(y)
\end{bmatrix}.
\end{align}
Accordingly, components of the constitutive equations can be written as:
\begin {align}
\hat\sigma_{xx} &=-C_L\kappa y+C_T\varepsilon_{yy}(y)+\zeta \mu_T E_{y,y}(y),\\
\hat\sigma_{yy} &=-C_T\kappa y+C_L\varepsilon_{yy}(y)+\zeta \mu_L E_{y,y}(y),\\
\hat\sigma_{xy} &=\hat\sigma_{yx} =0,\\~\nonumber\\
\widetilde\sigma_{yyy} &= l_1^2C_L\varepsilon_{yy,y}(y)- (1-\zeta) \mu_L E_{y}(y)-l_1^2C_T\kappa,\\
\widetilde\sigma_{xyx} &=\widetilde\sigma_{yxx} =- (1-\zeta) \mu_S E_{y}(y),\\
\widetilde\sigma_{xxy} &=l_1^2C_T\varepsilon_{yy,y}(y)- (1-\zeta) \mu_T E_{y}(y)-l_1^2C_L\kappa,\\
\widetilde\sigma_{xxx} &=\widetilde\sigma_{xyy}=\widetilde\sigma_{yyx}=\widetilde\sigma_{yxy}=0,\\~\nonumber\\
\hat{D}_y &=\epsilon E_y(y)+(1-\zeta) \mu_L\varepsilon_{yy,y}(y)-(1-\zeta)\mu_T\kappa ,\\
\hat{D}_x &=0 ,\\~\nonumber\\
\widetilde{D}_{xx} &=\zeta \mu_L\kappa y-\zeta \mu_T\varepsilon_{yy}(y),\\
\widetilde{D}_{yy} &=l_2^2\epsilon E_{y,y}(y)-\zeta \mu_L\varepsilon_{yy}(y)+\zeta \mu_T\kappa y,\\
\widetilde{D}_{xy} &=\widetilde{D}_{yx}=0.
\end{align}
Therefore, the components of physical stress and physical electric displacement are:
\begin{align}
\sigma_{xx}=&-C_L\kappa y+C_T\varepsilon_{yy}(y)+ \mu_T E_{y,y}(y)-l_1^2C_T\varepsilon_{yy,yy}(y),\\
\sigma_{yy}=&-C_T\kappa y+C_L\varepsilon_{yy}(y)+ \mu_L E_{y,y}(y)-l_1^2C_L\varepsilon_{yy,yy}(y),\label{SigmaYYBending}\\
\sigma_{xy}=&\sigma_{yx}=0,\\~\nonumber\\
D_{y}=& \epsilon E_y(y)+\mu_L\varepsilon_{yy,y}(y)-\mu_T\kappa-l_2^2\epsilon E_{y,yy}(y),\label{DYBending}\\
D_{x}=&0.
\end{align}
Therefore, equilibrium equations in $y$-direction read:
\begin{align}
&-C_T\kappa+C_L\varepsilon_{yy,y}(y)+ \mu_L E_{y,yy}(y)-l_1^2C_L\varepsilon_{yy,yyy}(y)=0,\label{equilibrium1bending}\\
& \epsilon E_{y,y}(y)+\mu_L\varepsilon_{yy,yy}(y)-l_2^2\epsilon E_{y,yyy}(y)=0\label{equilibrium2bending},
\end{align}
subjected to homogeneous Neumann boundary conditions (Eqs.~\eqref{SNBC2},\eqref{SNBC4}):
\begin{align}
& l_1^2C_L\varepsilon_{yy,y}(y)-(1-\zeta) \mu_L E_{y}(y)-l_1^2C_T\kappa=0,  \quad \text{on} \quad  y=\{-\frac{T}{2}, \frac{T}{2}\},\label{highorderNeumann1bending}\\
& l_2^2\epsilon E_{y,y}(y)-\zeta \mu_L\varepsilon_{yy}(y)+\zeta \mu_T\kappa y=0,   \quad \text{on} \quad   y=\{-\frac{T}{2}, \frac{T}{2}\}\label{highorderNeumann2bending}.
\end{align}
The low-order homogeneous Neumann boundary conditions (Eqs.~\eqref{SNBC1},\eqref{SNBC3}) will be satisfied with the solution we will postulate. 

Extracting $E_{y,y}$ and $\varepsilon_{yy,y}$ from \eqref{SigmaYYBending} and \eqref{DYBending}, respectively, and replacing them into \eqref{equilibrium1bending} and \eqref{equilibrium2bending}, respectively, we obtain:
\begin{align}
&(l_1^2l_2^2)E_{y,yyyy}(y)-(l_1^2+l_2^2+l_{\mu}^2)E_{y,yy}(y)+E_y(y)=\frac{D_y-l_1^2D_{y,yy}}{\epsilon}-\frac{\kappa}{\epsilon}(\mu_L\frac{C_T}{C_L}-\mu_T),\\
&(l_1^2l_2^2)\varepsilon_{yy,yyyy}(y)-(l_1^2+l_2^2+l_{\mu}^2)\varepsilon_{yy,yy}(y)+\varepsilon_{yy}(y)=\frac{\sigma_{yy}-l_2^2\sigma_{yy,yy}}{C_L}+\frac{C_T}{C_L}\kappa y.
\end{align}
where $l_{\mu}^2=\frac{\mu_L^2}{C_L\epsilon}$.

One can postulate the solutions to be of the form:
\begin{align}
E_y=&k_1\exp\left(\frac{-y-T/2}{a_1}\right)+k_2\exp\left(\frac{-y-T/2}{a_2}\right)+k_3\exp\left(\frac{y-T/2}{a_1}\right)+k_4\exp\left(\frac{y-T/2}{a_2}\right)-\frac{\kappa}{\epsilon}\left(\mu_L\frac{C_T}{C_L}-\mu_T\right),\label{electricfieldbending}\\
\varepsilon_{yy}(y)=&k_5\exp\left(\frac{-y-T/2}{a_1}\right)+k_6\exp\left(\frac{-y-T/2}{a_2}\right)+k_7\exp\left(\frac{y-T/2}{a_1}\right)+k_8\exp\left(\frac{y-T/2}{a_2}\right)+\frac{C_T}{C_L}\kappa y,\label{strainbending}
\end{align}
for $y\in [-T/2, T/2]$, with: 
\begin{align}
a_1,a_2 = \sqrt{\dfrac{2l_1^2l_2^2}{l_1^2+l_2^2+l_{\mu}^2\pm\sqrt{(l_1^2+l_2^2+l_{\mu}^2)^2-4l_1^2l_2^2}}}.
\end{align}
Note that $k_1=k_3$,  $k_2=k_4$, $k_5=-k_7$, $k_6=-k_8$ due to symmetry.
Substituting the $E_y$ and $\varepsilon_{yy}$ from Eqs.~\eqref{electricfieldbending} and \eqref{strainbending} to Eqs.~\eqref{equilibrium1bending}, \eqref{equilibrium2bending}, \eqref{highorderNeumann1bending} and \eqref{highorderNeumann2bending}, we obtain the unknowns $k_i$. 
\begin{align}
k_1=k_3=&\frac{-\kappa(C_L\mu_T - C_T\mu_L)(2a_2\hat\gamma_2(\zeta -1)  + \hat\beta_2T\zeta)}{2C_L\epsilon(a_1\hat\beta_2\hat\gamma_1 - a_2\hat\beta_1\hat\gamma_2)},\\
k_2=k_4=&\frac{ \kappa(C_L\mu_T - C_T\mu_L)(2a_1\hat\gamma_1(\zeta -1)  + \hat\beta_1T\zeta)}{2C_L\epsilon(a_1\hat\beta_2\hat\gamma_1 - a_2\hat\beta_1\hat\gamma_2)},\\
k_5=-k_7=&\frac{-a_1\alpha_1\kappa(C_L\mu_T - C_T\mu_L)(2a_2\hat\gamma_2(\zeta -1)+ \hat\beta_2T\zeta)}{2C_L\mu_L(a_1\hat\beta_2\hat\gamma_1 - a_2\hat\beta_1\hat\gamma_2)},\\
k_6=-k_8=&\frac{ a_2\alpha_2\kappa(C_L\mu_T - C_T\mu_L)(2a_1\hat\gamma_1(\zeta -1) + \hat\beta_1T\zeta)}{2C_L\mu_L(a_1\hat\beta_2\hat\gamma_1 - a_2\hat\beta_1\hat\gamma_2)}.
\end{align}
where
\begin{align}
\alpha_i=&1-\frac{l_2^2}{a_i^2},\\
\hat\beta_i=&(1+\exp(-T/a_i))\left(\frac{l_1^2}{l_\mu^2}\alpha_i+1-\zeta\right),\\
\hat\gamma_i=&(1-\exp(-T/a_i))\left(\frac{l_2^2}{a_i^2}(1-\zeta)+\zeta\right).
\end{align}

\bibliography{References}

\end{document}